\documentclass[aps,twocolumn]{revtex4}

\newcommand{\bq}{\begin{quotation}\noindent}
\newcommand{\eq}{\end{quotation}}
\newcommand{\bc}{\begin{center}}
\newcommand{\ec}{\end{center}}
\def\tr{{\rm tr}\,}
\def\drangle{\rangle\!\rangle}
\def\dlangle{\langle\!\langle}

\newtheorem{assump}{Assumption}
\newtheorem{defn}{Definition}

\usepackage{graphicx}
\usepackage{indentfirst}
\usepackage{amssymb}

\begin{document}

\title{A Quantum-Bayesian Route to Quantum-State Space}

\author{Christopher A. Fuchs$^\dagger$ and R\"udiger Schack$^\sharp$
\medskip
\\
\small
$^\dagger$Perimeter Institute for Theoretical Physics
\\
\small
Waterloo, Ontario N2L 2Y5, Canada
\medskip\\
\small
$^\sharp$Department of Mathematics, Royal Holloway, University of London
\\
\small
Egham, Surrey TW20 0EX, United Kingdom}

\date{25 September 2009}

\begin{abstract}
  In the quantum-Bayesian approach to quantum foundations, a quantum state is viewed as an expression
  of an agent's personalist Bayesian degrees of belief, or probabilities, concerning the results of measurements. These
  probabilities obey the usual probability rules as required by Dutch-book
  coherence, but quantum mechanics imposes additional constraints upon them. In this paper, we explore the question of
  deriving the structure of quantum-state space from a set of assumptions in
  the spirit of quantum Bayesianism. The starting point is the representation
  of quantum states induced by a symmetric informationally complete
  measurement or SIC. In this representation, the Born rule takes the form of
  a particularly simple modification of the law of total probability. We
  show how to derive key features of quantum-state space from (i) the
  requirement that the Born rule arises as a simple modification of the law of
  total probability and (ii) a limited number of additional assumptions of a
  strong Bayesian flavor.
\end{abstract}

\maketitle

\section{Introduction}
\label{sec:intro}

In the standard formulation of (finite-dimensional) quantum mechanics, a
quantum state is a density operator, $\rho$, on a $d$-dimensional
Hilbert space.  A measurement with $m$ outcomes is described by a POVM,
$\{E_1,\ldots,E_m\}$, a collection of positive semi-definite operators that sum to the
identity. The probability, $p(i)$, of the $i$-th measurement outcome
is given by the Born rule,
\begin{equation}  \label{eq:Born}
p(i)=\tr(\rho E_i) \;.
\end{equation}
If the POVM $\{E_i\}$ is informationally complete \cite{Caves02}, the
state $\rho$ is fully determined by the outcome probabilities $\{p(i)\}$. With
respect to some fiducial informationally complete POVM, the vector of
probabilities $p(i)$ is thus an alternative description of the quantum state.
This means that quantum-state space can be viewed as a subset of the
probability simplex.

According to the quantum-Bayesian approach to quantum foundations \cite{Caves02,Fuchs01,Schack01,Fuchs02,Fuchs03,Schack04,Fuchs04,Caves07,Appleby05a,Appleby05b,Timpson08,longPaper}, the probabilities $p(i)$ represent
an agent's Bayesian degrees of belief, or personalist probabilities
\cite{Ramsey26,DeFinetti31,Savage54,DeFinetti90,Bernardo94,Jeffrey04}. They are numbers expressing the agent's
uncertainty about which measurement outcome will occur and acquire an operational meaning
through decision theory \cite{Bernardo94}.  Quantum-Bayesian state assignments are personalist in the sense that they are
functions of the agent alone, not functions of the world external to the agent
\cite{Caves07}. In other words, there are---in principle---potentially as many quantum states for a given quantum system as there are agents who care to take note of it.  Nonetheless, despite not being specified by agent-independent facts, personalist probability assignments are far from arbitrary. Dutch-book coherence \cite{Ramsey26,DeFinetti90,Jeffrey04} as a normative principle requires that an agent's degrees of belief conform to the usual rules of the probability calculus, and this is a surprisingly powerful constraint when coupled with the agent's overall belief system \cite{Logue95}.

In addition to the rules required by Dutch-book coherence, the Born rule
(\ref{eq:Born}) puts further constraints on the probabilities used in quantum
mechanics. From this arises two questions which are of central importance for
the quantum-Bayesian program. One question, on which there has been much
progress recently \cite{longPaper,Appleby09a,Appleby09b}, is that of the mathematical structure of the
set of probabilities resulting from Eq.~(\ref{eq:Born}). The second question
concerns the origin of the quantum-mechanical constraints on the agent's
probabilities, i.e., the origin of the Born rule. The authors' present view on
this question is that the Born rule should be seen as an empirical addition to
Dutch-book coherence \cite{longPaper}.

What we mean by this is the following.  Dutch-book coherence, though a normative rule, is of a purely logical character \cite{Skyrms87}.  The way Bernardo and Smith \cite{Bernardo94} put its significance is this:
\bq
\small
Bayesian Statistics offers a rationalist theory of personalistic beliefs in contexts of uncertainty, with the central aim of characterising how an individual should act in order to avoid certain kinds of undesirable behavioural inconsistencies.  \ldots\   The goal, in effect, is to
establish rules and procedures for individuals concerned with disciplined uncertainty accounting.  The theory is not descriptive, in the sense of claiming to model actual behaviour.  Rather, it is prescriptive, in the sense of saying `if you wish to avoid the possibility of these undesirable consequences you must act in the following way.'
\eq
On the other hand, to a quantum Bayesian it is crucial that there is no such thing as a ``right and true'' quantum state \cite{Caves07}.  But if so, what is one to make of the Born rule in Eq.~(\ref{eq:Born})?  What are these things $\rho$ and $E_i$ that the probabilities are being calculated from?  The meaning of the rule calls for an explanation in our terms.  Our solution is to think of the Born rule in a normative way, rather than as a strict law of nature.  It is something along the lines, but not identical to, Dutch-book coherence:  The Born rule should be viewed as a normative principle for relating one's various degrees of belief about the outcomes of various measurements.  The idea is that if one does not make sure his probability assignments are related according to the dictum of the Born rule, nature is liable to give ``undesirable consequences'' for his decisions. In contrast to usual Dutch-book coherence, though, the origin of the normative rule is not of a purely logical character.  It should rather be seen as dependent upon contingent features of the particular physical world we happen to live in.

To shed further light on the similarities and differences between Dutch book
coherence and the Born rule, here we renew the question of deriving the
structure of quantum-state space from a set of assumptions formulated
and motivated fully in terms of the probability assignments of a Bayesian
agent. In Section \ref{sec:assumptions} we show one way to derive several key
features of quantum-state space from the assumption that the Born rule arises
as a simple modification of the law of total probability, complemented by
a few further assumptions of a strong Bayesian character.  Many details of the derivation are left out since a full account can be found in \cite{longPaper}.  Indeed the role of the present paper should be viewed as a supplement to the (very long) \cite{longPaper}. Here we try to be as logically crisp as possible, deleting extended examples and motivations, and very carefully labeling every assumption.

The set of assumptions in Section \ref{sec:assumptions} was inspired by the
observation that the mapping between the Hilbert space and probability simplex
formulations of quantum mechanics becomes very simple if the fiducial POVM is
a symmetric informationally complete POVM, or SIC \cite{Zauner99,Caves99,Renes04,Fuchs04b,Appleby05,ApplebyDangFuchs}.  In
particular, in this representation the Born rule takes the form of an
extremely simple modification of the law of total probability \cite{longPaper}, thus
motivating our main assumption in Section \ref{sec:assumptions}. This
motivation will be explained in Section \ref{sec:SIC}, where we prepare the
ground for the rest of the paper by reviewing
SICs briefly and showing how to rewrite the Born rule in terms of them.

\section{SICs and the Born rule}
\label{sec:SIC}

Consider a set of $d^2$ one-dimensional projection operators, $\Pi_i$, in
$d$-dimensional Hilbert space such that
\begin{equation} \label{eq:trPiPj}
\tr \,\Pi_i\Pi_j=\frac{d\delta_{ij}+1}{d+1}\;.
\end{equation}
The informationally complete POVM $\{E_i\}$ defined by $E_i=\frac1d\Pi_i$ is
called a SIC \cite{Caves99}. SICs have been explicitly proven to exist in dimensions
$d=2$--$15$, 19, and 24 (see references in \cite{GrasslScott}).  Furthermore, they have been observed by
computational means, to a numerical precision of $10^{-38}$, in dimensions
$d=2$--$67$ \cite{GrasslScott}.  For this paper, we will assume that SICs exist in
all dimensions.

With respect to a SIC, a density operator can be recovered easily from
the outcome probabilities given by the Born rule (\ref{eq:Born}) by using the
beautiful formula \cite{Caves99,Fuchs04b}
\begin{equation}  \label{eq:inverseBorn}
\rho = \sum_{i=1}^{d^2}\left( (d+1)p(i) - \frac1d \right)\Pi_i \;.
\end{equation}
Using this equation one can show that, in the representation
induced by a fiducial SIC, the set of all quantum states can be characterized
very elegantly. According to one such characterization, a
probability vector $p(i)$ is a pure quantum state if and only if it satisfies
the constraints \cite{ApplebyDangFuchs}
\begin{equation}
\sum_i p(i)^2=\frac{2}{d(d+1)}
\label{PurePurity}
\end{equation}
and
\begin{equation}
\sum_{ijk} c_{ijk}\,p(i)p(j)p(k)=\frac{d+7}{(d+1)^3}\;,
\label{ChocolateMouse}
\end{equation}
where the coefficients $c_{ijk}$ are defined by
\begin{equation}
c_{ijk}=\mbox{Re}\;\tr\!\Big(\Pi_i\Pi_j\Pi_k\Big)\;.
\label{TripleSec}
\end{equation}
All other quantum states, which means all mixed states, are constructed by
taking convex combinations of the states given by Eqs.~(\ref{PurePurity}) and
(\ref{ChocolateMouse}). A key question for the quantum Bayesian program is how
to understand and motivate the structure of quantum-state space expressed
in these equations as restrictions on an agent's personalist probability
assignments.

\begin{figure}
\includegraphics[height=3.3in]{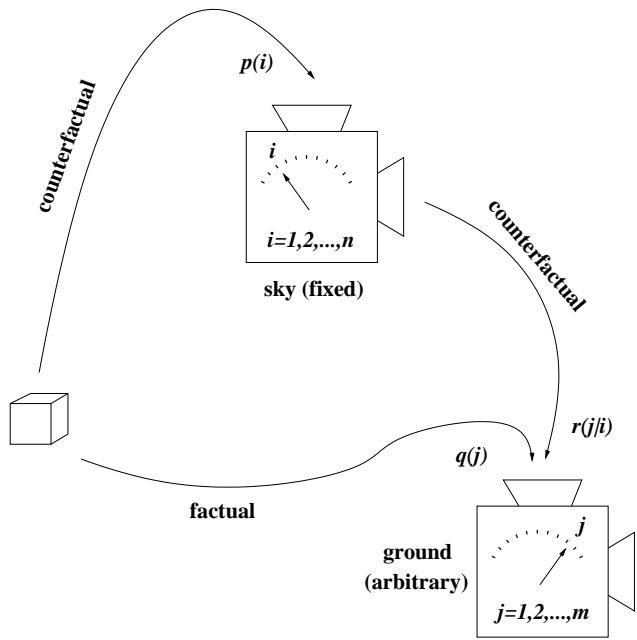}
\bigskip\caption{This diagram shows the conceptual framework of this paper.  A system is imagined to be measurable in two ways.
  The measurement on the ground, with outcomes $j=1,\ldots,m$, is an arbitrary
  measurement that could be performed in the laboratory. The measurement in
  the sky, with outcomes $i=1,\ldots,n$, is a fixed fiducial measurement
  introduced to analyze the measurement on the ground. The probability
  distributions $p(i)$ and $r(j|i)$ represent an agent's probabilities
  assuming the measurement in the sky is actually performed. The probability
  distribution $q(j)$ represents instead the agent's probabilities under the
  assumption that the measurement in the sky is {\it not\/} performed. If the
  measurement in the sky is a SIC with $n=d^2$ outcomes, $q(j)$ is related to
  $p(i)$ and $r(j|i)$ by the simple relation (\ref{eq:urgleichung}).}
\end{figure}

A strong hint as to where to look for an answer is given by the surprising form
the Born rule takes when written in SIC language.
Consider the scenario in Figure 1, where one measurement (which we call the
``measurement on the ground'') is analyzed in terms of another measurement
(the ``measurement in the sky''), and assume for the time being that the
measurement in the sky is a SIC, implying that it has $n=d^2$ outcomes. The
probabilities $p(i)$ are thus a representation of the agent's prior state
assignment. The conditional probabilities $r(j|i)$ are the agent's
probabilities to obtain outcome $j$ on the ground assuming that the
measurement in the sky was actually performed and resulted in outcome $i$.

Let us now denote by $s(j)$ the agent's probabilities for the outcomes of the
measurement on the ground in this situation, i.e., when the measurement in the
sky was actually carried out before the measurement on the ground. Dutch-book
coherence requires that the agent computes $s(j)$ from $p(i)$ and
$r(j|i)$ by using the law of total probability,
\begin{equation} \label{eq:totalProb}
s(j)=\sum_{i=1}^{d^2} p(i) r(j|i) \;.
\end{equation}

If, however, the measurement in the sky is not performed and remains
counterfactual, Dutch-book coherence places no constraints on the agent's
probabilities, denoted by $q(j)$, for the outcomes on the ground. In this
case, it is the quantum formalism through the Born rule that restricts the
agent's distribution $q(j)$. This follows easily by setting $q(j)=\tr \rho
F_j$ and $r(j|i)=\tr \Pi_i F_j$ for some general POVM $\{F_j\}$ with $m$
outcomes and using Eq.~(\ref{eq:inverseBorn}). One obtains
\begin{equation}
q(j)=\sum_{i=1}^{d^2} \left((d+1)p(i)-\frac1d\right) r(j|i) \;.
\label{eq:urgleichung}
\end{equation}
This is the Born rule $q(j)=\tr\rho F_j$ expressed in the SIC
representation. As in Ref.~\cite{longPaper} we stress the central importance
of this equation by calling it the {\it Ur\-glei\-chung}. The similarity with the
law of total probability is striking---it is basically the simplest modification of that law that it could possibly be \cite{Ferrie09}. To get from the law of total probability to the Ur\-glei\-chung, one merely makes the replacement
\begin{equation}
\label{eq:affineTrafo}
\sum_i p(i) r(j|i) \longrightarrow \sum_i
f\big(p(i)\big) r(j|i) \;,
\end{equation}
with $f$ simply an affine mapping, $f(x)=(d+1)x-\frac1d$. The functional form of the Born rule
expressed by the Ur\-glei\-chung will be the pivot for the development in the
next section.

\section{Deriving the structure of quantum-state space}
\label{sec:assumptions}

\setcounter{assump}{0}
\setcounter{defn}{0}

In this section, the main section of the paper, we formulate a series of
assumptions from which a number of key features of the structure of quantum-state space can be derived. As stated previously we omit many of the mechanical details of the proofs for greater clarity.  The omitted details can be found in \cite{longPaper}. The basic situation we consider is as in Figure 1. We imagine a fiducial, counterfactual
$n$-outcome measurement ``in the sky'' in terms of which we analyze an
arbitrary $m$-outcome measurement ``on the ground.'' Initially there are no
restrictions on the numbers $n$ and $m$.  As before, $p(i)$, the {\it prior in
  the sky}, represents the probabilities in the sky, and $q(j)$ represents the
probabilities on the ground.  We write $r(j|i)$ to represent the conditional
probability for obtaining $j$ on the ground, given that $i$ was found in the
sky.  When we want to suppress components, we will write vectors $\|p\drangle$
and $\|q\drangle$, and write $R$ for the matrix with entries $r(j|i)$---by
definition, $R$ is a stochastic matrix.

We start by postulating a generalized Ur\-glei\-chung where we take the mapping
$f$ in Eq.~(\ref{eq:affineTrafo}) to be a general affine mapping $f(x)=\alpha
x-\beta$.

\begin{assump}{\rm $\!\!$:}   \label{ass:urgleichung}
{\rm Generalized Ur\-glei\-chung}.
For any measurement on the ground, $q(j)$ should be calculated according to
\begin{equation}
q(j)= \sum_{i=1}^n \big(\alpha p(i)-\beta\big) r(j|i) \;,
\label{BigBoy}
\end{equation}
where $\alpha$ and $\beta$ are fixed nonnegative real numbers.
\end{assump}
Since the $q(j)$ are probabilities, they satisfy the double inequality $0\le
q(j)\le1$ or
\begin{equation}
0\le\sum_{i=1}^n \big(\alpha p(i)-\beta\big) r(j|i) \le1 \;.
\label{eq:urungleichung}
\end{equation}
We call this double inequality the {\it Ur\-un\-glei\-chung}. It puts immediate
restrictions on the distributions $p(i)$ and $r(j|i)$, i.e., on the vector
$\|p\drangle$ and the matrix $R$. For an agent to accept quantum mechanics
it means, at least in part, he commits to these restrictions on his Bayesian probability assignments.
Our ultimate goal---which in this paper we will achieve only partially---is
the precise characterization of these restrictions in Bayesian terms. We will
denote by $\mathcal P_0$ the set of all priors for the sky permitted by
quantum mechanics, and by $\mathcal R_0$ the set of all permitted conditional
distributions $R$. Our next assumption is about the sets $\mathcal P_0$ and
$\mathcal R_0$. To formulate it, we need a definition.

\begin{defn}{\rm $\!\!$:} Let $\mathcal P$ be a set of priors in the sky and
  let $\mathcal R$ be a set of stochastic matrices. We say that $\mathcal P$
  and $\mathcal R$ are {\rm consistent} if all pairs $\big(\|p\drangle,R\big)
  \in {\mathcal P}\times {\mathcal R}$ obey the Ur\-un\-glei\-chung
  (\ref{eq:urungleichung}). Furthermore, we say $\mathcal P$ and $\mathcal R$
  are {\rm maximal} whenever ${\mathcal P}^\prime\supseteq {\mathcal P}$ and
  ${\mathcal R}^\prime \supseteq {\mathcal R}$ imply ${\mathcal
    P}^\prime={\mathcal P}$ and ${\mathcal R}^\prime={\mathcal R}$ for any
  consistent ${\mathcal P}^\prime$ and ${\mathcal R}^\prime$.
\end{defn}

\begin{assump}{\rm $\!\!$:} \label{ass:maximality}
{\rm Maximality:} The sets $\mathcal P_0$ and
  $\mathcal R_0$ of all valid priors for the sky and all valid conditionals
  $R$ are taken to be consistent and maximal.
\end{assump}
In other words, we assume that quantum mechanics restricts the set of
probabilities available to the agent as little as possible given the universal
validity of the generalized Ur\-glei\-chung (\ref{BigBoy}).

Unfortunately, Assumption~\ref{ass:maximality} does not fix the sets $\mathcal
P_0$ and $\mathcal R_0$ uniquely. There are many consistent and maximal pairs
$(\mathcal P,\mathcal R)$. The assumptions below constitute one way to proceed
toward the goal of a complete characterization of $\mathcal P_0$ and $\mathcal
R_0$. There is little doubt that there exist simpler and more compelling sets
of assumptions to achieve this goal. Finding these is work in progress.

\begin{assump}{\rm $\!\!$:} \label{ass:ignorance}
{\rm Possibility of complete ignorance:} The
  constant vector
 \begin{equation}  \label{eq:ignorance}
\|p\drangle=
\left(\frac{1}{n},\frac{1}{n},\ldots,\frac{1}{n}\right)^{\!\rm
    T}
\end{equation}
 is in the set $\mathcal P_0$.
\end{assump}
This assumption makes sure that the agent can be in a state of complete
ignorance about the outcome of the measurement in the sky.

\begin{assump}{\rm $\!\!$:} \label{ass:span}
{\rm Priors span the simplex:} The elements of
  $\mathcal P$ span the probability simplex in $n$ dimensions.
\end{assump}
If this assumption were not satisfied, one could use a smaller simplex
for all considerations.

\begin{assump}{\rm $\!\!$:} \label{ass:reciprocity}
{\rm Principle of Reciprocity:}  For any $R\in\mathcal R_0$ and any
outcome $j$ on the ground, the vector $\|p\drangle$ with components
\begin{equation}
p(i)=\frac{r(j|i)}{\sum_k r(j|k)}
\label{eq:reciprocity}
\end{equation}
is in the set $\mathcal P_0$ of valid priors for the sky. Conversely, all valid
priors $\|p\drangle\in\mathcal P_0$ can be written in this way.
\end{assump}
To motivate this assumption and its name, imagine that both the measurement in
the sky and the measurement on the ground are performed and the agent learns
the outcome $j$ on the ground while remaining ignorant of the outcome in the
sky. Imagine further that his prior in the sky before the measurement
is given by the state (\ref{eq:ignorance}) of complete ignorance. The
expression (\ref{eq:reciprocity}) is then the agent's posterior probability
for the outcome $i$ in the sky, given the outcome $j$ on the ground, as
computed by Bayes's rule. The content of the Principle of Reciprocity is
that the set of priors in the sky is equal to the set of posteriors upon learning the outcome on the ground.

The assumptions so far are very natural and already lead to a number of
interesting consequences \cite{longPaper}. For instance, it follows immediately
from Assumption~\ref{ass:urgleichung}
that the relation
\begin{equation}  \label{eq:nmab}
\alpha = n\beta +1
\end{equation}
holds between the three constants of the generalized Ur\-glei\-chung
(\ref{BigBoy}). Assumption~\ref{ass:maximality} implies that the sets
$\mathcal P_0$ and $\mathcal R_0$ are both convex, and even compact \cite{Appleby09b}, so that they necessarily have well-defined extreme points. And
Assumption~\ref{ass:reciprocity} implies the existence of an important class
of special priors:

\begin{defn}{\rm $\!\!$:}
Let the measurement on the ground be identical to the measurement
  in the sky. Denote the components of the matrix $R$ by $r_s(j|i)$ in this
  case. By the Principle of Reciprocity (Assumption~\ref{ass:reciprocity}),
  the distributions $\|e_k\drangle$, $k=1,\ldots,n$, with components
\begin{equation}    \label{eq:basis}
e_k(i)=\frac{r_s(k|i)}{\sum_l r_s(k|l)}
\end{equation}
are in the set $\mathcal P_0$. They are called {\rm basis states}.
\end{defn}
Using Assumption~\ref{ass:span}, one can show that
these components take the form
\begin{equation}    \label{eq:basisComponents}
e_k(i)=\frac{1}{\alpha}(\delta_{ki}+\beta)
\end{equation}
and satisfy the relation
\begin{equation}    \label{eq:basisConstraints}
\sum_i e_k(i)^2=\frac{1}{\alpha^2}\Big(1+2\beta+n\beta^2\Big)\;.
\end{equation}

However, to pin down the sets $\mathcal P_0$ and $\mathcal R_0$
further, and in particular to fix the parameterized form of the
constants $n$, $\alpha$, and $\beta$---i.e., that $n=d^2$,
$\alpha=d+1$, and $\beta=\frac1d$---we need two additional
postulates. First here is another definition.

\begin{defn}{\rm $\!\!$:} A measurement on the ground is said to have the
  property of {\rm in-step unpredictability} (ISU) if a uniform prior in the
  sky implies a uniform probability assignment for the probabilities on the
  ground, i.e., for an ISU measurement, whenever $\|p\drangle$ is the uniform
  distribution (\ref{eq:ignorance}), then $\|q\drangle$ is given by the
  uniform distribution
  $\left(\frac{1}{m},\frac{1}{m},\ldots,\frac{1}{m}\right)^{\!\rm T}$.
\end{defn}
The existence of ISU measurements, which will be postulated in Assumption
\ref{ass:certainty} below, means that an agent may be totally ignorant about
both the (counterfactual) outcome in the sky and the outcome on the ground.

Let us now denote by $r_{\rm \scriptscriptstyle ISU}(j|i)$ the components of the
matrix $R$ for a measurement on the ground with $m$ outcomes, $m\ne n$, and
in-step unpredictability. It can be shown \cite{longPaper} that one must have
\begin{equation}
\sum_i r_{\rm \scriptscriptstyle ISU}(j|i) = \frac{n}{m}\;.
\end{equation}
By the Principle of Reciprocity, this ISU measurement gives rise to a class of
priors which we denote by $\|p_k\drangle$, $k=1,\ldots,m$. Their components
are given by
\begin{equation}   \label{eq:ISUprior}
p_k(i)=\frac{m}{n} r_{\rm \scriptscriptstyle ISU}(k|i) \;;
\end{equation}
each vector $\|p_k\drangle$ represents a valid prior in the sky. This leads to
our next definition.

\begin{defn}{\rm $\!\!$:}
  We say that a measurement with in-step unpredictability {\rm achieves the
    ideal of certainty} if $\|p\drangle=\|p_k\drangle$ implies that
  $q(j)=\delta_{jk}$, i.e., for such a measurement and a prior in the sky
  given by $\|p_k\drangle$, the agent is certain that the outcome on the
  ground will be $k$.
\end{defn}
This is a very specific definition. It is motivated by the following
consideration. Consider a setup with an ISU measurement on the ground, i.e., a
measurement with total ignorance for both ground and sky, and imagine the
measurement in the sky is actually performed. Observing $k$ on the ground
while remaining ignorant about the sky then gives rise to the posterior
$\|p_k\drangle$ for the sky (see the discussion following Assumption
\ref{ass:reciprocity}). Now go back to the usual situation in which the
measurement in the sky remains counterfactual, and assume the agent's prior
for the sky is $\|p_k\drangle$. If the measurement achieves the ideal of
certainty, the agent will be certain that the measurement on the ground
results in the very outcome $k$.

\begin{assump}{\rm $\!\!$:}
\label{ass:certainty} {\rm Availability of Certainty.}  For any system, there is
a measurement with in-step unpredictability of some number $m_0\ge2$ of outcomes
that (i) achieves the ideal of certainty and (ii) for which one of the priors
$\|p_k\drangle$ defined in Eq.~(\ref{eq:ISUprior}) has the form of a basis
distribution (\ref{eq:basis}).
\end{assump}
For a measurement of this type, we have that \cite{longPaper}
\begin{equation}
\dlangle p_j\|
p_k\drangle=\frac{1}{\alpha}\left(\frac{m_0}{n}\delta_{jk}+\beta\right)\;,
\;\;\;\;j,k=1,\ldots,m_0\;,
\label{Delapidate}
\end{equation}
where $\dlangle \cdot\| \cdot\drangle$ denotes the inner
product. Using condition (ii) of the above assumption, it follows that
the squared norm $\dlangle p_k\|p_k\drangle$ of any of the vectors
$\|p_k\drangle$ is equal to the squared norm of the basis vectors given
by Eq.~(\ref{eq:basisConstraints}). This, together with
Eq.~(\ref{eq:nmab}) now implies the equality
\begin{equation}
\frac{m_0}{n}\alpha-\beta=1
\label{NimbleNooph}
\end{equation}
for any measurement satisfying Assumption~\ref{ass:certainty}.

Equation~(\ref{Delapidate}) expresses that any two of the vectors
$\|p_k\drangle$ differ by the same angle, $\theta$, defined by
\begin{equation}
\cos\theta = \frac{\dlangle p_1\|p_2\drangle}{\dlangle p_1\|p_1\drangle}\;.
\end{equation}
Using the relations~(\ref{eq:nmab})
and~(\ref{NimbleNooph}) between our four variables, $\alpha$, $\beta$, $n$ and
$m_0$ established above, this angle can be seen to equal
\begin{equation}    \label{eq:gral}
\cos\theta = \frac{n-m_0}{(m_0-1)^2 + n-1} \;.
\end{equation}
We are now ready to state our last assumption.

\begin{assump}{\rm $\!\!$:} \label{ass:QBibbo} {\rm Many Systems, Universal Angle.}
The identity of a system is parameterized by its pair $(n,m_0)$.  Nonetheless for all
systems, the angle $\theta$ between pairs of priors $\|p_k\drangle$ for any measurement satisfying
Assumption~\ref{ass:certainty} is a universal constant given by $\cos\theta=1/2$.
\end{assump}
The value $\cos\theta=1/2$ is less arbitrary than it may appear at first
sight. Taken by itself, the assumption that $\theta$ is universal implies
that, for any $m_0\ge2$, there is an integer $n$ such that the right-hand side of
Eq.~(\ref{eq:gral}) evaluates to the constant $\cos\theta$. It is not hard to
show that this is possible only if this constant is of the form
\begin{equation}
\cos\theta = \frac q{q+2} \;,
\end{equation}
where $q$ is a non-negative integer. The universal angle postulated above
corresponds to the choice $q=2$.

Every choice for $q$ leads to a different relation between $n$ and $m_0$. For
$q=0$, we find $n=m_0$, in which case the Urgleichung turns out to be identical
to the classical law of total probability. For $q=1$, we get the relationship
$n=\frac12 m_0(m_0+1)$ which, although this fact plays no role in our
argument, is characteristic of theories defined in real Hilbert space \cite{Wootters86}.
And for $q=2$, we obtain
\begin{equation}   \label{eq:square}
n=m_0^2 \;.
\end{equation}

Equations (\ref{NimbleNooph}) and (\ref{eq:square}) hold for the special
measurement postulated in Assumption~\ref{ass:certainty}. If we eliminate
$m_0$ from these equations we find, with the help of Eq.~(\ref{eq:nmab}), the
relationships
\begin{equation}
n = (\alpha-1)^2\;, \;\;\; \beta=\frac1{\sqrt n}\;.
\end{equation}
These equalities form the main result of this paper. They must hold for
{\it any\/} measurement on the ground. If we denote the integer $\alpha-1$ by
the letter $d$, we  recover the constants of the original
Ur\-glei\-chung (\ref{eq:urgleichung}).

With the Ur\-glei\-chung in the form (\ref{eq:urgleichung}) as the starting
point and minimal additional assumptions, a large amount of detailed
information about the structure of quantum-state space can be derived. Details
can be found in Ref.~\cite{longPaper} and the paper by Appleby, Ericsson, and
Fuchs \cite{Appleby09b} in this special issue.

\section{Summary and conclusion}

Our main postulate, the generalized Ur\-glei\-chung or
Assumption~\ref{ass:urgleichung}, is an addition to standard Dutch-book
coherence. It restricts an agent's probability assignments in a situation
involving a counterfactual measurement---the measurement in the sky---where
Dutch-book coherence does not impose any specific constraints. The form of the
generalized Ur\-glei\-chung is given by a minimal modification of the law of total
probability, which is the law connecting the agent's probabilities in the case
the measurement in the sky is {\it factualized},
i.e., actually carried out. This means that the key assumption of this paper
arises through a formal connection between an agent's probabilities in two
complementary scenarios, one in which the measurement in the sky remains
counterfactual and one in which it is factualized.

Assumption~\ref{ass:maximality} guarantees that the set of probability
assignments available to the agent is maximal within the constraints set by
the generalized Ur\-glei\-chung, i.e., Assumption~\ref{ass:maximality} makes sure
that the agent's probability assignments are not unduly restricted. In a
similar spirit, Assumption~\ref{ass:ignorance} guarantees that the state of
complete ignorance is among the agent's potential priors, and
Assumption~\ref{ass:span} makes sure that the set of priors available to the
agent is large enough to span the probability simplex.

With Assumption~\ref{ass:reciprocity}, the Principle of Reciprocity, we return
to the theme of exploring connections between the two respective scenarios of
a counterfactual and a factualized measurement in the sky. The Principle of
Reciprocity states that the set of priors for the sky available to the agent
should be identical to the set of the agent's posteriors for a factualized
measurement in the sky. The question of what motivates the particular relation
between probabilities for counterfactual and factualizable measurements
expressed in Assumptions~\ref{ass:urgleichung} and~\ref{ass:reciprocity}
strikes us as a mysterious and important one.

The numerical relation between the constants $\alpha$, $\beta$, and $n$, and
in particular the fact that $n$ is a perfect square, follows from the
existence of a single special measurement defined in
Assumption~\ref{ass:certainty}, together with the postulate of a universal
angle in Assumption~\ref{ass:QBibbo}. These last two assumptions, as well as
the first five, are given purely in terms of the personalist probabilities a
Bayesian agent may assign to the outcomes of certain experiments. Nowhere in
all this do we mention amplitudes, Hilbert space, or any other part of the
usual apparatus of quantum mechanics. What has been sketched in this paper
constitutes a novel approach to the quantum formalism, providing fresh insight
for the foundations of quantum mechanics.  Maybe even more importantly, the
success of this approach provides a compelling case for quantum Bayesianism
\cite{Caves02,Fuchs01,Schack01,Fuchs02,Fuchs03,Schack04,Fuchs04,Caves07,Appleby05a,Appleby05b,Timpson08,longPaper}.

\section{Acknowledgements}

CAF thanks Wayne Myrvold for discussions on the logical structure of coherence arguments, and especially Lucien Hardy for discussions on the ``signature'' $q$ of a theory.  This research was supported in part by the U.~S. Office of Naval Research (Grant No.\ N00014-09-1-0247). Research at Perimeter Institute is supported by the Government of Canada through Industry Canada and by the Province of Ontario through the Ministry of Research \& Innovation.

\end{document}